\documentclass[letter, traditabstract]{aa} 
\usepackage{graphicx}
\usepackage{natbib}
\bibpunct{(}{)}{,}{a}{}{;}
\usepackage{amssymb}
\usepackage{amsmath}
\usepackage{url}
\usepackage{multirow}
\usepackage{txfonts}
\begin{document}
   \title{Water in low-mass star-forming regions with \textit{Herschel}\thanks{\textit{Herschel} is an ESA space observatory with science instruments provided by European-led Principal Investigator consortia and with important participation from NASA.} (WISH-LM)}
\subtitle{High-velocity H$_\textrm{2}$O bullets in L1448-MM observed with HIFI}
\titlerunning{Water in low-mass star-forming regions with \textit{Herschel} (WISH-LM)}


\author{L.E.~Kristensen\inst{1}
\and E.F. van Dishoeck\inst{1,2} 
\and M. Tafalla\inst{3} 
\and R. Bachiller\inst{3} 
\and B. Nisini\inst{4} 
\and R. Liseau\inst{5} 
\and U.A. Y{\i}ld{\i}z\inst{1}
}

\institute{
Leiden Observatory, Leiden University, PO Box 9513, 2300 RA Leiden, The Netherlands \and
Max Planck Institut f\"{u}r Extraterrestrische Physik, Giessenbachstrasse 1, 85748 Garching, Germany \and
Observatorio Astron\'{o}mico Nacional (IGN), Calle Alfonso XII,3. 28014, Madrid, Spain \and
INAF - Osservatorio Astronomico di Roma, 00040 Monte Porzio catone, Italy \and
Department of Earth and Space Sciences, Chalmers University of Technology, Onsala Space Observatory, 439 92 Onsala, Sweden
}

\date{Draft: \today}


\abstract
{\textit{Herschel}-HIFI observations of water in the low-mass star-forming object L1448-MM, known for its prominent outflow, are presented, as obtained within the `Water in star-forming regions with \textit{Herschel}' (WISH) key programme. Six H$_2^{16}$O lines are targeted and detected ($E_{\rm up}$/$k_{\rm B}$$\sim$50--250 K), as is CO $J$= 10--9 ($E_{\rm up}/k_{\rm B}$$\sim$305 K), and tentatively H$_2^{18}$O 1$_{10}$--1$_{01}$ at 548 GHz. All lines show strong emission in the ``bullets'' at $|\varv|$$>$50 km s$^{-1}$ from the source velocity, in addition to a broad, central component and narrow absorption. The bullets are seen much more prominently in H$_2$O than in CO with respect to the central component, and show little variation with excitation in H$_2$O profile shape. Excitation conditions in the bullets derived from CO lines imply a temperature $>$150 K and density $>$10$^5$ cm$^{-3}$, similar to that of the broad component. The H$_2$O/CO abundance ratio is similar in the ``bullets'' and the broad component, $\sim$0.05--1.0, in spite of their different origins in the molecular jet and the interaction between the outflow and the envelope. The high H$_2$O abundance indicates that the bullets are H$_2$ rich. The H$_2$O cooling in the ``bullets'' and the broad component is similar and higher than the CO cooling in the same components. These data illustrate the power of \textit{Herschel}-HIFI to disentangle different dynamical components in low-mass star-forming objects and determine their excitation and chemical conditions.}


\maketitle

\section{Introduction}

Low-mass star formation is accompanied by the launch of powerful jets that drive strong shocks into the parental material \citep[e.g.,][]{arce07}, both in the form of shell shocks that interact with the molecular envelope, and in the form of the jet itself \citep[e.g.,][]{hirano10}. The jets are generally not uniform, but consist of small condensations (``bullets''\footnote{Bullets and extremely high-velocity (EHV) gas are used interchangeably throughout.}), that can be attributed to internal working surfaces caused by episodic ejection of material from the protostar \citep{santiago-garcia09}. The bullets are chemically rich, in particular in oxygen-bearing species \citep{tafalla10}, and should therefore display high abundances of water. H$_2$O is one of the best shock tracers because the abundance can increase by orders of magnitude up to $\lesssim$10$^{-4}$ through sputtering of grain mantles and formation in the gas phase at high temperatures \citep{flower10}. Also, due to the large velocity gradients in shocks, emission is effectively optically thinner than in a quiescent object \citep{franklin08}.

H$_2$O has previously been observed in outflowing gas from young stellar objects (YSOs) using, e.g., \textit{Odin}, SWAS, and ISO. The Heterodyne Instrument for the Far-Infrared (HIFI) on \textit{Herschel} opens up spectrally resolved, high angular resolution observations of both H$_2$O and high-$J$ CO in low-mass protostars \citep[e.g.,][]{lefloch10, yildiz10, kristensen10}. The latter paper presents H$_2$O observations of the low-mass star-forming region NGC1333. The H$_2$O line profiles are complex, consisting of three velocity components characterised by their width. The broadest component ($\Delta\varv$$>$20 km s$^{-1}$) originates in the interaction between the outflow and the envelope; the medium-broad component (5$<$$\Delta\varv$$<$20 km s$^{-1}$) arises where the jet impacts on the inner, dense envelope; a narrow absorption component ($\Delta\varv$$<$5 km s$^{-1}$) is attributed to the cold outer envelope. These initial HIFI results show that H$_2$O is one of the best dynamics tracers in YSOs, and confirm that the H$_2$O/CO abundance increases with velocity to $\sim$1. To date, H$_2$O has not been uniquely identified in the bullets, and its abundance with respect to other jet species is unknown. Moreover, the different nature of the shocks could lead to different H$_2$O abundances.

L1448-MM is a low-mass, Class 0 protostar located in Perseus \citep[$L$=11.6 $L_\odot$, $M_{\rm env}$=1.9 $M_\odot$, $d$ = 235 pc;][]{vandishoeck11}. This object is the prototype of the class of sources showing bullet emission in CO \citep{bachiller90}, with red- and blue-shifted emission peaks at $\sim$ $\pm$50 km s$^{-1}$ with respect to $\varv_{\rm LSR}$ = $+$5.2 km s$^{-1}$. The bullets are clearly visible in other shock tracers such as SiO \citep[e.g.][]{nisini07}, and have been imaged at high spatial resolution using interferometers \citep[e.g.,][]{guilloteau92, hirano10}. Our observations of H$_2$O within the framework of the `Water in star-forming regions with \textit{Herschel}' \citep[WISH;][]{vandishoeck11} key programme further explore the chemical and excitation conditions of shocked gas in low-mass star-forming regions.

\section{Observations and results}
\label{sec:obs}

The central position of L1448-MM (03$^{\rm h}$25$^{\rm m}$38\fs9; $+$30\degr44\arcmin05\farcs4; J2000) was observed with HIFI on \textit{Herschel} \citep{degraauw10, pilbratt10} in seven different settings covering six H$_2^{16}$O, two H$_2^{18}$O and one CO transition ($E_{\rm u}/k_{\rm B}\approx50-300$ K; Table \ref{tab:h2o_line} available online). Data were obtained using the dual beam-switch mode with a nod of 3\arcmin, except for the ground-state ortho-H$_2$O line at 557 GHz, where a position switch to ($+$10\arcmin; $+$5\arcmin) was used. The diffraction-limited beam size ranges from 19\arcsec\ to 39\arcsec\ (4500--9500 AU). Data were reduced using HIPE ver. 4.0. The calibration uncertainty is taken to be 10\% for lines observed in Bands 1, 2, and 5 while it is 30\% in Band 4. The pointing accuracy is $\sim$2\arcsec. A main-beam efficiency of 0.65--0.75 is adopted (Table \ref{tab:h2o_line}). Subsequent analysis of the data is performed in CLASS including subtraction of linear baselines. H- and V-polarizations are co-added after inspection; no significant differences are found between the two data sets. $^{12}$CO 3--2 and $^{13}$CO 3--2 were observed with the James Clerk Maxwell Telescope (JCMT). CO 2--1 was observed at the IRAM 30m \citep{tafalla10}, and CO 4--3 at the JCMT \citep{nisini00}.

\begin{figure}
\begin{center}
\includegraphics[width=0.9\columnwidth, angle=0]{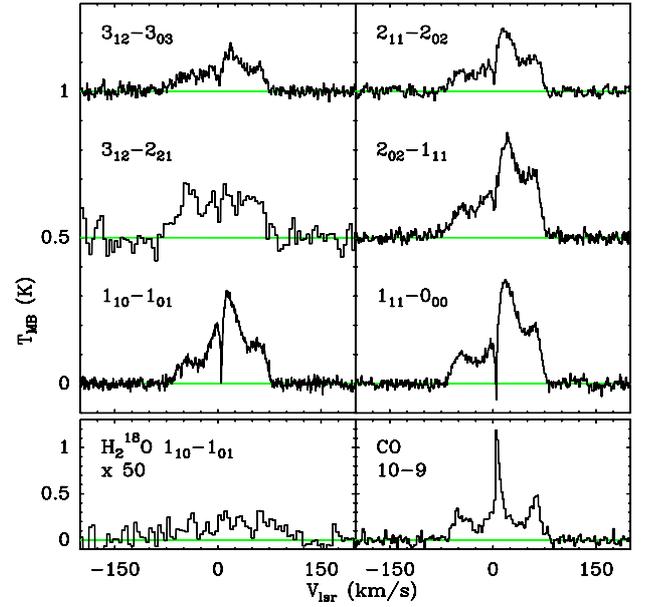}
\end{center}
\caption{Continuum-subtracted HIFI H$_2$O, H$_2^{18}$O and CO spectra obtained at the central position of L1448-MM ($\varv_{\rm source}$ = 5.2 km s$^{-1}$).}
\label{fig:spectra}
\end{figure}

All targeted lines are detected, except H$_2^{18}$O 1$_{11}$--0$_{00}$, observed with the main isotopologue (Fig. \ref{fig:spectra}). This figure strikingly demonstrates that the bullets are much more prominent in H$_2$O than in CO relative to the central component. Also, the H$_2$O profiles are very different from those seen in NGC 1333 \citep{kristensen10}. Each line consists of multiple components: two Gaussian peaks at velocities of $\sim$~$\pm$50~km~s$^{-1}$ with respect to $\varv_{\rm source}$; a broad ($\sim$50 km s$^{-1}$) component centered close to $\varv_{\rm source}$; and a narrow ($\sim$5 km s$^{-1}$) component seen in absorption at $\varv_{\rm source}$. The extremely high-$\varv$ emission (henceforth EHV-B and EHV-R) is detected in all lines, although the detection in the H$_2^{18}$O 1$_{10}$--1$_{01}$ line is tentative (2$\sigma$; 0.13 K km s$^{-1}$). The emission components are well reproduced by Gaussians and fitted to obtain the integrated intensity (Table \ref{tab:h2o_obs}). The H$_2$O line profiles vary little with excitation. The shape of the EHV component in the CO profiles resembles that of H$_2$O (Fig. \ref{fig:gaussfit}, appendix), whereas the the broad component is much weaker in CO. The CO lines also contain a narrower component ($\sim$10 km s$^{-1}$) not seen in H$_2$O.

To compare observations done with different beam sizes, all observations are scaled to a common beam of 22\arcsec, the beam at 988 GHz. The scaling follows the recipe outlined in Appendix B of \citet{tafalla10}, where a linear scaling with beam size is found to be appropriate for the EHV components and a power-law scaling with index 0.75 for the broad component. It is only significant for the 557 GHz lines due to the larger beam (39\arcsec).

\begin{table}
\caption{H$_2$O, H$_2^{18}$O and CO emission in each velocity component\tablefootmark{a}.}
\scriptsize
\begin{center}
\begin{tabular}{r r r r r}
\hline \hline
           &      & EHV-B & EHV-R & Broad \\
Transition & rms\tablefootmark{b}  & $\int T_{\rm MB}~{\rm d}\varv$ & $\int T_{\rm MB}~{\rm d}\varv$ & $\int T_{\rm MB}~{\rm d}\varv$ \\
           & (mK) & (K\,km\,s$^{-1}$) & (K\,km\,s$^{-1}$) & (K\,km\,s$^{-1}$) \\
\hline
H$_2$O      1$_{10}$--1$_{01}$ &  9 &    2.54 &  3.00 &  14.4 \\
            1$_{11}$--0$_{00}$ & 11 &    3.98 &  4.00 &  14.8 \\
            2$_{02}$--1$_{11}$ & 14 &    4.01 &  3.86 &  17.6 \\
            2$_{11}$--2$_{02}$ & 14 &    2.47 &  2.53 &   9.9 \\
            3$_{12}$--3$_{03}$ & 13 &    2.34 &  1.14 &   7.8 \\
            3$_{12}$--2$_{21}$ & 66 &    6.22 &  3.32 &   7.9 \\
H$_2^{18}$O 1$_{10}$--1$_{01}$\tablefootmark{c} &  3 &     0.13 &     0.11 &     0.23 \\
            1$_{11}$--0$_{00}$\tablefootmark{d} & 11 &  $<$0.25 &  $<$0.19 &  $<$0.28 \\
CO 2--1                        & 51 &    2.79 &  4.12 &  38.5 \\
 3--2                          & 39 &    4.82 &  6.58 &  44.9 \\
 4--3                          & 160 &   11.9 &  15.1 &  74.6 \\
10--9                          & 66 &    7.30 &  8.42 &  17.8 \\
\multicolumn{2}{l}{FWHM\tablefootmark{e} (km\,s$^{-1}$)} & 38.4$\pm$3.2 & 22.4$\pm$3.0 &  47.6$\pm$4.5 \\
\multicolumn{2}{l}{$\varv_{\rm LSR}$\tablefootmark{e} (km\,s$^{-1}$)} & $-$43.4$\pm$1.3 & 59.3$\pm$1.8 &  16.8$\pm$2.2 \\ \hline
\end{tabular}
\tablefoot{
	\tablefoottext{a}{Obtained from Gaussian fits to each component. All Gauss-fit parameters are provided in Table \ref{tab:all_obs} available online.}
	\tablefoottext{b}{Measured in 1 km\,s$^{-1}$ bins.}
	\tablefoottext{c}{Obtained by fixing the position and width of Gaussian functions and only fitting the intensity.}
	\tablefoottext{d}{Upper limits are 3$\sigma$.}
	\tablefoottext{e}{Intensity-weighted average of values determined from Gaussian fits to the H$_2$O lines. Uncertainties include statistical errors only.}}
\end{center}
\label{tab:h2o_obs}
\end{table}

The tentative detection of the H$_2^{18}$O 1$_{10}$--1$_{01}$ (548 GHz) line is used to constrain the optical depth, $\tau$, of the main isotopologue, assuming that the line itself is optically thin and that [$^{16}$O]/[$^{18}$O]=550. The optical depth, $\tau$, is $\sim$25 in each EHV component and 9 in the broad component implying that emission from H$_2^{16}$O 1$_{10}$--1$_{01}$ is optically thick. The 3$\sigma$ upper limit on the H$_2^{18}$O 1$_{11}$--0$_{00}$ line implies $\tau$$\lesssim$10 and $\lesssim$30 in H$_2^{16}$O for the broad and EHV components, respectively. $^{13}$CO 3--2 emission is not detected in any of the components. The $^{12}$CO 3--2 optical depth is $\lesssim$2 and $\lesssim$0.3 in the EHV and broad components, respectively, implying that the CO emission is optically thin.

\begin{figure}
\begin{center}
\includegraphics[width=0.95\columnwidth, angle=0]{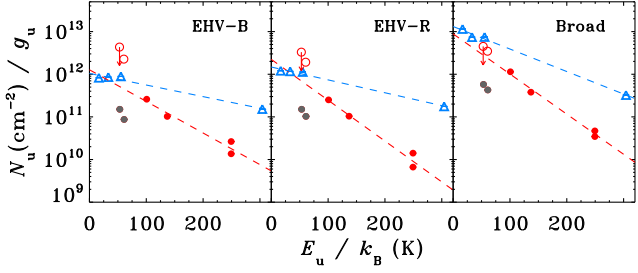}
\end{center}
\caption{Rotational diagrams of the EHV and broad components for H$_2$O (red circles) and CO (blue triangles). CO has been shifted downwards by a factor of 100. The ground-state transitions (grey) are excluded from the fit (dashed lines). H$_2^{18}$O intensities $\times$550 are shown as open circles.}
\label{fig:rotdiag}
\end{figure}

The integrated intensities of each of the EHV and broad components are used to construct Boltzmann diagrams, where the column density of each upper level in the optically thin limit ($N_{\rm up}^*$) per sub-level is plotted versus $E_{\rm up}$ (Fig. \ref{fig:rotdiag}). In the diagrams, $N_{\rm up}^*$ for the H$_2$O ground-state transitions is lower than expected, confirming that the true $N_{\rm up}$ is higher and that the lines are optically thick. Excluding these two lines from further analysis, the resulting rotational temperature is remarkably similar for all three components ($\sim$45--60 K), while the beam-averaged column density of the broad component is $\sim$5 times higher than that of the EHV components ($\sim$10 and 2$\times$10$^{13}$ cm$^{-2}$, respectively; Table \ref{tab:rotdiag}). Rotational diagrams are also made for CO and results are given in Table \ref{tab:rotdiag}. $T_{\rm rot}$ is significantly higher, 150 K in the EHV components and 80 K in the broad. The column density is also higher, $\sim$10$^{16}$ cm$^{-2}$, and the inferred H$_2$O/CO ratio is $\sim$0.002.

\begin{table}
\caption{Rotational-diagram results for each velocity component.}
\scriptsize
\begin{center}
\begin{tabular}{l c c c c c}
\hline \hline
          & \multicolumn{2}{c}{H$_2$O} && \multicolumn{2}{c}{CO} \\  \cline{2-3} \cline{5-6}
Component & $T_{\rm rot}$ &                   $N$ && $T_{\rm rot}$ & $N$ \\
          &           (K) & (10$^{13}$ cm$^{-2}$) &&           (K) & (10$^{15}$ cm$^{-2}$) \\ \hline
EHV-B     &      58$\pm$4 &           1.9$\pm$0.4 &&     160$\pm$18 & \phantom{2}6.0$\pm$0.7 \\
EHV-R     &      45$\pm$3 &           2.4$\pm$0.5 &&     144$\pm$14 & \phantom{2}7.6$\pm$0.8 \\
Broad     &      46$\pm$2 &           9.5$\pm$2.1 &&      82$\pm$4 & 38.6$\pm$3.9 \\ \hline
\end{tabular}
\end{center}
\label{tab:rotdiag}
\end{table}

\section{Discussion}
\label{sec:disc}

\subsection{Excitation conditions}

To determine the physical parameters, a small grid of RADEX slab models was run \citep{vandertak07}. This grid is first used to determine CO excitation conditions by fitting $T_{\rm rot}$ (see Fig. \ref{fig:radex_rotdiag} in the on-line appendix). To reproduce $T_{\rm rot}$$\sim$150 K as observed in the EHV components, a kinetic temperature of 150 K and density greater than $\sim$10$^6$ cm$^{-3}$ is required. Alternatively, a kinetic temperature of $\sim$500 K and density of $\sim$10$^5$ cm$^{-3}$ can also reproduce the observed CO rotational temperature. In the following, both sets of values, ($T$, $n$) = (150 K, 10$^6$ cm$^{-3}$; model 1), and (500 K, 10$^5$ cm$^{-3}$; model 2), will be examined to constrain the H$_2$O excitation. Similar values are needed to reproduce the lower CO rotational temperature of the broad component, $T_{\rm rot}$$\sim$80 K, ($T$, $n$) = (100 K, 10$^6$ cm$^{-3}$) and (500 K, 5$\times$10$^4$ cm$^{-3}$). The latter solution is excluded, however, because the envelope density within the 20$\arcsec$ beam is always $>$ 4$\times$10$^5$ cm$^{-3}$. Independent analysis of outflow emission \citep{nisini99, nisini00} indicate that H$_2$O is excited in a hotter, denser medium than CO, and for this reason a model with ($T$, $n$) = (500 K, 10$^6$ cm$^{-3}$; model 3), is included in the analysis.

To constrain the H$_2$O excitation, the same physical conditions as determined for CO are taken as fixed and line widths of 40 and 20 km s$^{-1}$ for the EHV-B and EHV-R components, respectively, and 50 km s$^{-1}$ for the broad component are adopted. C-type shock models show that CO and H$_2$O cooling takes place at the same location in a shock \citep{flower10}, thus validating this assumption. The H$_2$O o/p ratio is assumed to be 3, and the only variable in the RADEX modelling is the total H$_2$O column density. To constrain this, the ratio of the higher-excited H$_2^{16}$O lines is taken with respect to the 2$_{02}$--1$_{11}$ line, and the calculated emission in this line is constrained to be greater than the observed value. The same is done for the H$_2^{18}$O tentative detection and upper limit. A $\chi^2$ analysis is used to determine the best-fit value of the H$_2$O column density. Once the column density is determined, the predicted intensity is compared to the observed intensity of the 2$_{02}$--1$_{11}$ line, and from this the filling factor is determined. The predicted column density is re-scaled to the average column density in the beam, and compared to $N$(CO) to determine the H$_2$O/CO abundance ratio. 

\begin{figure}
\begin{center}
\includegraphics[width=0.85\columnwidth, angle=0]{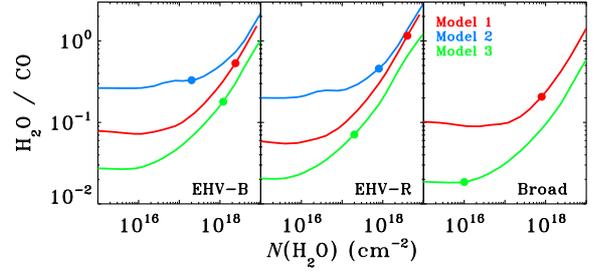}
\end{center}
\caption{H$_2$O/CO ratio as a function of H$_2$O column density for the three different models 1, 2 and 3. The best-fit results are marked with points.}
\label{fig:chi2}
\end{figure}

The best-fit results are listed in Table \ref{tab:radex}. $N$(H$_2$O) is found to be $\sim$10$^{16}$--10$^{18}$ cm$^{-2}$ for all components and all three models, implying a beam filling factor of $\lesssim$0.01--0.1. This leads to an average H$_2$O/CO abundance ratio of $\sim$0.05--1.0 in all of the components; the range is caused by the excitation model rather than differences between the physical components. This result is very different from the rotational-diagram analysis above, where the abundance ratio is a factor of $>$100 lower. The difference arises because H$_2$O is not thermalised and optically thick, as opposed to CO, and thus non-LTE, radiative-transfer models are needed to treat the excitation properly. Figure \ref{fig:chi2} shows the H$_2$O/CO ratio as function of $N$(H$_2$O) demonstrating that only model 3 produces abundance ratios $\lesssim$0.1. 

The RADEX models that match the H$_2$O line emission best predict that the filling factor for the EHV component is very small, $<$0.01. Such low values agree with high spatial-resolution observations showing this component to be confined to the molecular jet and consisting of several sub-arcsec knots \citep[e.g.,][]{hirano10}. The model furthermore predicts that the ground-state transitions have optical depths in excess of 100 (Fig. \ref{fig:radex_tau}, online appendix), and that the two H$_2^{18}$O transitions and the higher-excited H$_2^{16}$O transitions are also optically thick ($\tau$$\sim$3--4). Thus, the optical depth derived from the line ratios can only be considered a relative optical depth between the two transitions. If the H$_2^{18}$O 1$_{10}$--1$_{01}$ transition is assumed to be optically thin, a typical H$_2$O column density of a few times 10$^{15}$ cm$^{-2}$ is required. However, models with this column density predict that the intensity of the 2$_{02}$--1$_{11}$ line is close to that observed, or lower in the case of the broad component, implying filling factors of unity inconsistent with the bullet nature. The CO emission remains optically thin ($\tau$$<$1) for all three models even with the small filling factors derived for H$_2$O, consistent with the non-detection of $^{13}$CO 3--2.

\begin{table}
\caption{Modelled H$_2$O line emission and inferred parameters.}
\scriptsize
\begin{center}
\begin{tabular}{l c c c c c}
\hline \hline
 & $N$(H$_2$O) & $\chi^2_{\rm red}$ & $\int$ $T$(2$_{02}$--1$_{11}$) d$\varv$ & Filling factor & H$_2$O/CO \\
 & (10$^{18}$ cm$^{-2}$)\tablefootmark{a} & & (K km s$^{-1}$) & ($\times$10$^{-3}$) & \\ \hline
Model 1\tablefootmark{b}:\\
EHV-B & 2.4 & 0.4 &           3000 & 1.3 & 0.5 \\
EHB-R & 0.2 & 0.4 & \phantom{2}870 & 4.4 & 0.1 \\
Broad & 1.2 & 1.5 &           2070 & 8.3 & 0.3 \\
Model 2\tablefootmark{c}:\\
EHV-B & 4.0 & 0.9 &           2650 &           1.5 & 1.0 \\
EHB-R & 0.8 & 0.4 & \phantom{2}890 &           4.3 & 0.5 \\
Model 3\tablefootmark{d}:\\
EHV-B & 0.8\phantom{0} & 0.5 &           3800 &           1.1 & 0.1 \\
EHB-R & 0.08           & 0.9 & \phantom{3}890 &           4.3 & 0.05 \\
Broad & 0.01           & 0.2 & \phantom{3}290 & 65\phantom{.0} & 0.02 \\ \hline
\end{tabular}
\tablefoot{
	\tablefoottext{a}{Ortho- $+$ para-H$_2$O.}
	\tablefoottext{b}{($T$, $n$) = (150 K, 10$^6$ cm$^{-3}$) and (100 K, 10$^6$ cm$^{-3}$) for the EHV and broad components, respectively.}
	\tablefoottext{c}{($T$, $n$) = (500 K, 10$^5$ cm$^{-3}$).}
	\tablefoottext{d}{($T$, $n$) = (500 K, 10$^6$ cm$^{-3}$) for all components.}
}
\end{center}
\label{tab:radex}
\end{table}

\subsection{Physical origin}\label{sec:origin}

\subsubsection{EHV versus broad components}

Following the interpretation of the EHV emission in IRAS04191 \citep{santiago-garcia09} it seems likely that the knots are internal working surfaces (IWS) along the jet that appear as strongly beam-diluted emission in single-dish observations. Regardless of the precise shock details, the high abundance of H$_2$O in the bullets implies a high H$_2$ fraction, since otherwise reactions with atomic H drive H$_2$O back to OH and atomic O.

The timescale for H$_2$O to be produced in gas as a function of temperature has been determined by \citet{bergin98} for $n$=10$^5$ cm$^{-3}$. For $T$=500 K, the timescale is $\sim$100 years, which may be compared to the maximum dynamical time for a bullet with velocity 50 km s$^{-1}$ to leave a beam with radius 11\arcsec, 240 years. Thus, it is possible to reform H$_2$O from atomic gas over the timescales associated with the bullets. In the lower-$T$ scenario, H$_2$O will not form in the quantities observed here implying that the temperature is high in the bullets or that H$_2$O is not formed in the gas phase.

The gas-phase production of H$_2$O is in agreement with models of continuous spherical wind models of \citet{glassgold91}, where the formation takes place from a dust-free atomic gas in sufficient quantities if the mass-loss rate exceeds 10$^{-5}$ $M_\odot$ yr$^{-1}$, which is to be compared to the time-averaged mass-loss rate in the jet of $\sim$10$^{-6}$ $M_\odot$ yr$^{-1}$ \citep{hirano10}. The mass loss rate traced directly by the bullets is episodic (and thus higher than the time-averaged value) and the wind is not spherical, i.e., it is plausible that the H$_2$O abundance is consistent with production in the IWS from atomic gas. To resolve the issue of the origin of the molecules and to break the degeneracy between the different models, it is necessary to obtain velocity-resolved data of chemically related species, such as O and OH. Furthermore, measuring the H/H$_2$ ratio would test the models.

The broad component, centered near $\varv_{\rm LSR}$, is now observed in H$_2$O emission in a number of low-mass sources \citep[][in prep.]{kristensen10}. It is interpreted as arising in the interaction between the outflow and envelope, and is seen so prominently in H$_2$O emission (as opposed to, e.g., CO) due to the high production rate of H$_2$O at this location. H$_2$O is formed through both a series of neutral-neutral reactions pushing atomic oxygen into water, and through sputtering of icy grain mantles. Quantifying the contributions from the different mechanisms require observations of other grain-surface products, e.g., CH$_3$OH, and velocity-resolved observations of the intermediate gas-phase products O and OH. \citet{tafalla10} detect CH$_3$OH emission in both the EHV and broad components (but at an outflow position) indicating that a significant fraction of H$_2$O is sputtered from grains.

\subsubsection{Cooling}

The observed cooling rate in the three components is found by summing up the observed line emission (Table \ref{tab:cooling}). For the EHV components, the H$_2$O cooling rate is $\sim$ 3.5$\times$10$^{-4}$ $L_\odot$, while it is $\sim$1.3$\times$10$^{-3}$ $L_\odot$ for the broad. Extrapolating the cooling to include all lines with wavelengths greater than 60 $\mu$m gives a total H$_2$O luminosity of 2--4$\times$10$^{-2}$ $L_\odot$ depending on which model is used (Table \ref{tab:cooling}), i.e., the cooling by the lines observed with HIFI amounts to $\sim$10\%. The cooling is nearly the same for all three components demonstrating the power of spectrally resolving line profiles. This number may be compared to 4.5$\times$10$^{-2}$ $L_\odot$ as was found for H$_2$O by ISO-LWS in an 80\arcsec\ beam \citep{nisini00}, which applies to the sum of the components. 

The observed CO cooling rate is somewhat higher, $\sim$2$\times$10$^{-3}$ and 6.0$\times$10$^{-3}$ $L_\odot$, respectively. The CO cooling rate for the three components is extrapolated using the rotational diagrams to a total of 4.0$\times$10$^{-3}$ $L_\odot$, lower than the H$_2$O cooling. The EHV components contribute $\sim$50\% to the  CO cooling. ISO-LWS observed a total CO cooling rate of 3.0$\times$10$^{-2}$ $L_\odot$ \citep{nisini00}, but in a much larger beam encompassing the extended outflow emission and other physical components \citep{vankempen10}.

\begin{table}
\caption{H$_2$O and CO cooling rates in 10$^{-3}$ $L_\odot$.}
\scriptsize
\begin{center}
\begin{tabular}{l l c c c c}
\hline \hline
 & & EHV-B & EHV-R & Broad & Total \\ \hline
H$_2$O: &
Observed emission       & \phantom{0}0.4 & 0.3 & \phantom{0}1.1 & \phantom{0}1.8 \\
& Model 1 extrapol. & 11.0 & 4.8 & 26.6 & 42.4 \\
& Model 2 extrapol. & 10.7 & 6.4 &          \ldots & 17.1 \\
& Model 3 extrapol. & 15.1 & 6.5 & 11.7 & 33.3 \\
CO: &
Observed emission            & \phantom{0}0.2 & 0.2 & \phantom{0}0.4 & \phantom{0}0.8 \\
& $T_{\rm ex}$ extrapol. & \phantom{0}1.1 & 1.1 & \phantom{0}1.8 & \phantom{0}4.0 \\ \hline
\end{tabular}
\end{center}
\label{tab:cooling}
\end{table}

\section{Conclusions}

H$_2$O emission readily traces the EHV gas that has previously only been seen in very deep integrations of other species, such as CO. The EHV components are attributed to shocks in the molecular jet, while the underlying broad component, now seen in several low-mass YSOs, is associated with the interaction of the outflow and the envelope on larger scales. Thus, HIFI and H$_2$O together are ideal for revealing the dynamical components of low-mass star-forming regions. Both components are H$_2$ rich, since otherwise the H$_2$O abundance would be lower. The two distinct components appear remarkably similar in terms of excitation and chemical conditions in spite of their different chemistries. They have equal contributions from H$_2$O and CO cooling with H$_2$O being the dominant coolant of the two. High spectral-resolution observations are a prerequisite for interpreting spectrally unresolved PACS data of the same source.

\bibliographystyle{aa}
\bibliography{bibliography}

\begin{acknowledgements}
HIFI has been designed and built by a consortium of institutes and university departments from across Europe, Canada and the US under the leadership of SRON Netherlands Institute for Space Research, Groningen, The Netherlands with major contributions from Germany, France and the US. Consortium members are: Canada: CSA, U.Waterloo; France: CESR, LAB, LERMA, IRAM; Germany: KOSMA, MPIfR, MPS; Ireland, NUI Maynooth; Italy: ASI, IFSI-INAF, Arcetri-INAF; Netherlands: SRON, TUD; Poland: CAMK, CBK; Spain: Observatorio Astronomico Nacional (IGN), Centro de Astrobiolog{\ ́i}a (CSIC-INTA); Sweden: Chalmers University of Technology - MC2, RSS \& GARD, Onsala Space Observatory, Swedish National Space Board, Stockholm University - Stockholm Observatory; Switzerland: ETH Z{\"u}rich, FHNW; USA: Caltech, JPL, NHSC. Astrochemistry in Leiden is supported by NOVA, by a Spinoza grant and grant 614.001.008 from NWO, and by EU FP7 grant 238258. The WISH team is thanked for stimulating discussions.
\end{acknowledgements}

\Online

\begin{appendix}

\section{Gauss fit of data}

The data are fitted by Gaussians, with the exception of the narrow absorption component. Figure \ref{fig:gaussfit} shows the example of the H$_2$O 1$_{11}$--0$_{00}$ line, with the residual plotted below. Furthermore, Fig. \ref{fig:gaussfit} shows the CO 3--2 line scaled and overplotted on the H$_2$O 1$_{11}$--0$_{00}$ line, highlighting the similar line profile in the EHV components.

\begin{figure}
\begin{center}
\includegraphics[width=0.7\columnwidth, angle=0]{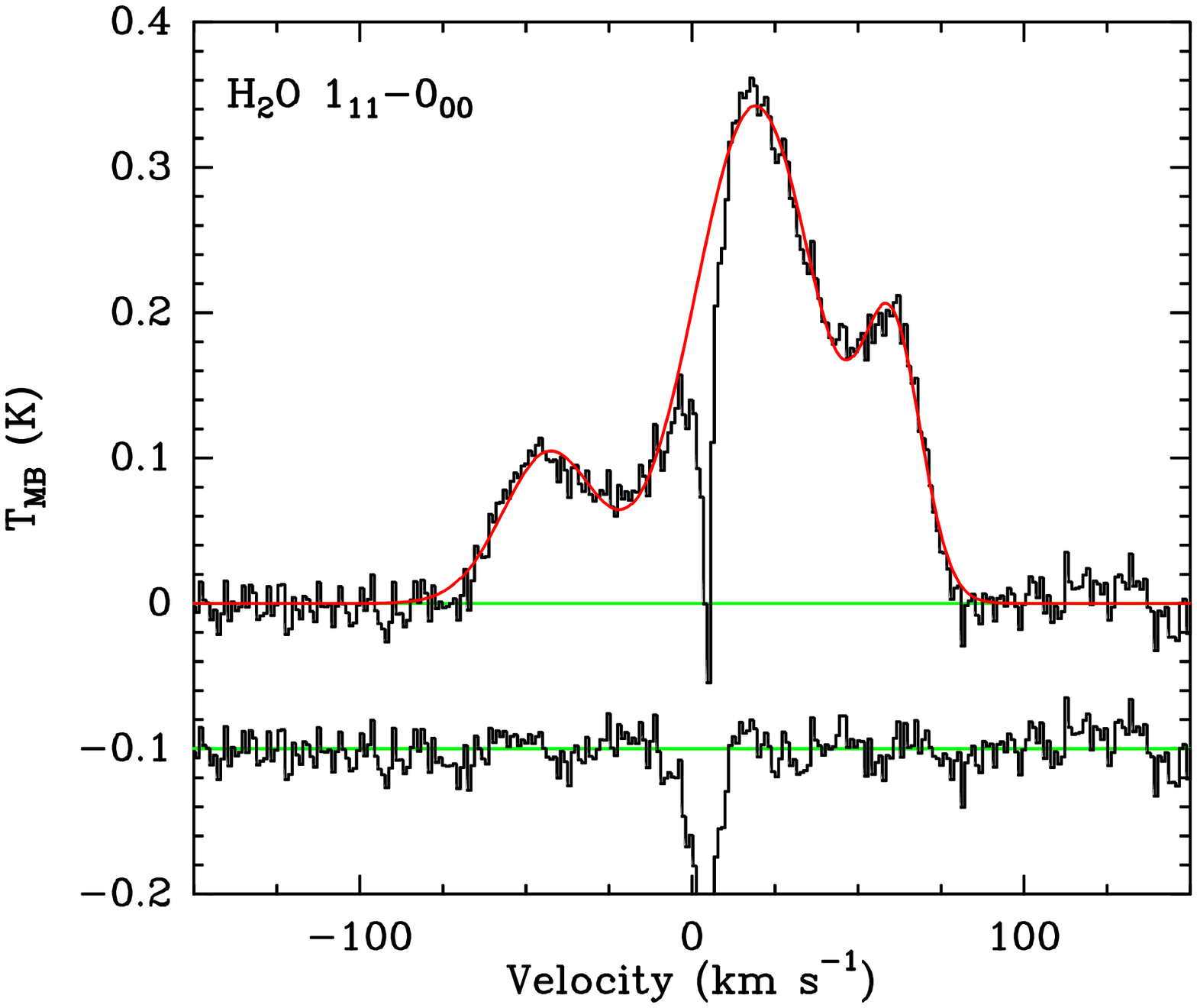}
\includegraphics[width=0.7\columnwidth, angle=0]{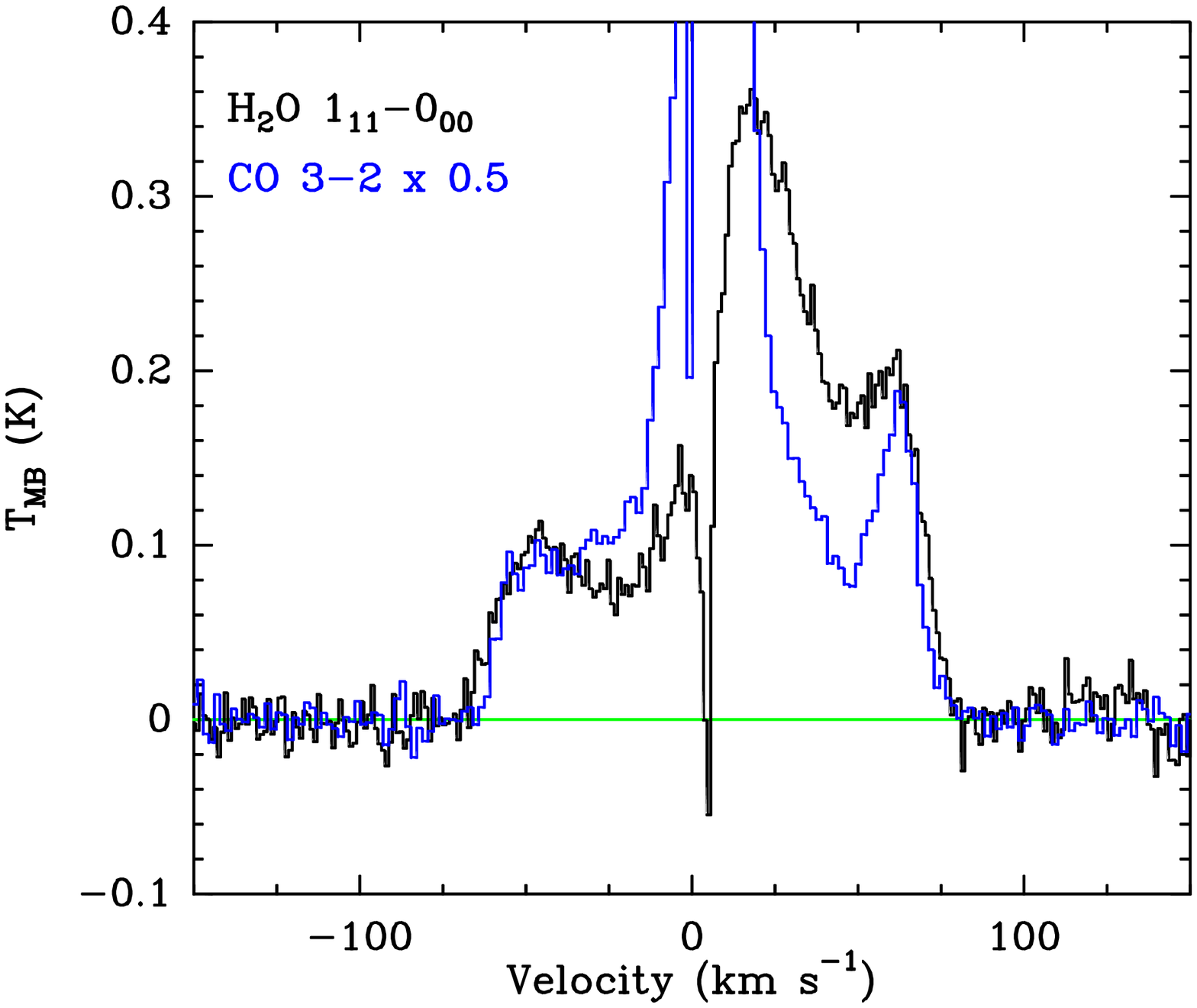}
\end{center}
\caption{(Top:) Gauss fit of the H$_2$O 1$_{11}$--0$_{00}$ line (red) with the residual shown below (offset at $-$0.1 K). The central absorption feature is not fitted. (Bottom:) CO 3--2 scaled to the peak intensity of H$_2$O in the bullets and overplotted on the H$_2$O 1$_{11}$--0$_{00}$ line.}
\label{fig:gaussfit}
\end{figure}

\onltab{2}{
\begin{table*}
\caption{H$_2$O, H$_2^{18}$O and CO emission in each velocity component at the source position\tablefootmark{a}.}
\scriptsize
\begin{center}
\begin{tabular}{r r r r r r c r r r r c r r r r}
\hline \hline
           &      & \multicolumn{4}{c}{EHV-B} && \multicolumn{4}{c}{EHV-R} && \multicolumn{4}{c}{Broad} \\
  \cline{3-6} \cline{8-11} \cline{13-16}
Transition & rms\tablefootmark{b}  & $T_{\rm MB}^{\rm peak}$ & $\int T_{\rm MB}~{\rm d}\varv$ & $FWHM$ & $\varv_{\rm LSR}$ && $T_{\rm MB}^{\rm peak}$ & $\int T_{\rm MB}~{\rm d}\varv$ & $FWHM$ & $\varv_{\rm LSR}$ && $T_{\rm MB}^{\rm peak}$ & $\int T_{\rm MB}~{\rm d}\varv$ & $FWHM$ & $\varv_{\rm LSR}$ \\
           & (mK) & (mK) & (K\,km\,s$^{-1}$) & (km s$^{-1}$) & (km s$^{-1}$) && (mK) & (K\,km\,s$^{-1}$) & (km s$^{-1}$) & (km s$^{-1}$) && (mK) & (K\,km\,s$^{-1}$) & (km s$^{-1}$) & (km s$^{-1}$) \\
\hline
H$_2$O 1$_{10}$--1$_{01}$ & 9 &  74 & 2.54 & 32.3 & $-$44.2 && 122 & 3.00 & 23.0 & 60.1 && 294 & 14.40 & 46.0 & 14.0 \\
      1$_{11}$--0$_{00}$ & 11 & 101 & 3.98 & 37.1 & $-$41.1 && 176 & 4.00 & 21.3 & 60.1 && 337 & 14.76 & 41.2 & 20.1 \\
      2$_{02}$--1$_{11}$ & 14 &  93 & 4.01 & 40.3 & $-$43.7 && 175 & 3.86 & 20.7 & 59.8 && 328 & 17.59 & 50.4 & 18.0 \\
      2$_{11}$--2$_{02}$ & 14 &  65 & 2.47 & 35.9 & $-$44.2 && 109 & 2.53 & 21.9 & 60.0 && 207 &  9.91 & 44.9 & 15.4 \\
      3$_{12}$--3$_{03}$ & 13 &  50 & 2.34 & 44.0 & $-$45.4 &&  65 & 1.14 & 16.5 & 60.7 && 124 &  7.76 & 54.5 & 16.3 \\
      3$_{12}$--2$_{21}$ & 66 & 147 & 6.22 & 39.7 & $-$43.1 && 111 & 3.32 & 28.1 & 55.4 && 139 &  7.94 & 53.4 & 14.8 \\
H$_2^{18}$O 1$_{10}$--1$_{01}$\tablefootmark{c} &  3 &      4 &    0.13 & \ldots & \ldots && 5 &  0.11 & \ldots & \ldots && 5 &  0.23 & \ldots & \ldots \\
            1$_{11}$--0$_{00}$\tablefootmark{d} & 11 & \ldots & $<$0.25 & \ldots & \ldots && \ldots & $<$0.19 & \ldots & \ldots && \ldots & $<$0.28 & \ldots & \ldots \\
CO 2--1 &  51 & 184 & 2.79 & 14.2 & $-$51.6 && 280 & 4.12 & 13.8 & 60.3 && 498 & 38.53 & 72.8 &  0.8 \\
   3--2 &  39 & 216 & 4.82 & 21.0 & $-$48.2 && 467 & 6.58 & 13.2 & 61.9 && 657 & 44.85 & 64.1 &  8.0 \\
   4--3 & 160 & 467 & 11.9 & 24.0 & $-$45.1 && 808 & 15.1 & 17.5 & 65.7 && 1023 & 74.61 & 68.5 & 19.0 \\ 
  10--9 &  66 & 269 & 7.30 & 25.5 & $-$47.2 && 426 & 8.42 & 18.6 & 61.3 && 323 & 17.77 & 51.6 & 10.1 \\
\hline
\end{tabular}
\tablefoot{
	\tablefoottext{a}{Obtained from Gaussian fits to each component.}
	\tablefoottext{b}{Measured in 1 km\,s$^{-1}$ bins.}
	\tablefoottext{c}{Obtained by fixing the position and width of Gaussian functions to the average value and only fitting the intensity.}
	\tablefoottext{d}{Upper limits are 3$\sigma$.}
}
\end{center}
\label{tab:all_obs}
\end{table*}
}

\onltab{3}{
\begin{table*}
\caption{H$_2$O, H$_2^{18}$O and CO transitions observed with \textit{Herschel}-HIFI\tablefootmark{a}.}
\scriptsize
\begin{center}
\begin{tabular}{r c c c c c c c}
\hline \hline
\multicolumn{1}{c}{Transition} & $\nu$ & $\lambda$   & $E_{\rm u}/k_{\rm B}$ & $A$  & Beam & $t_{\rm int}$\tablefootmark{b} & $\eta_{\rm MB}$ \\
                               & (GHz) & ($\mu$m)    & (K)   & ($10^{-3}$ s$^{-1}$) & ($''$) & (min.) \\
\hline
H$_2$O ~~~1$_{10}$--1$_{01}$ & \phantom{1}556.94 & 538.29 & \phantom{1}61.0 & \phantom{1}3.46 & 39 & 13.0 & 0.74 \\
1$_{11}$--0$_{00}$           &           1113.34 & 269.27 & \phantom{1}53.4 &           18.42 & 19 & 43.5 & 0.75 \\
2$_{02}$--1$_{11}$           & \phantom{1}987.93 & 303.46 & 100.8           & \phantom{1}5.84 & 22 & 23.3 & 0.73 \\
2$_{11}$--2$_{02}$           & \phantom{1}752.03 & 398.64 & 136.9           & \phantom{1}7.06 & 29 & 18.4 & 0.74 \\
3$_{12}$--3$_{03}$           &           1097.37 & 273.19 & 249.4           &           16.48 & 20 & 32.4 & 0.75 \\
3$_{12}$--2$_{21}$           &           1153.13 & 259.98 & 249.4           & \phantom{1}2.63 & 19 & 13.0 & 0.65 \\
\hline
H$_2^{18}$O ~~~1$_{10}$--1$_{01}$\phantom{\tablefootmark{c}} & \phantom{1}547.68 & 547.39 & \phantom{1}60.5 & \phantom{1}3.59 & 39 & 64.3 & 0.74 \\
1$_{11}$--0$_{00}$\tablefootmark{c} &           1101.70 & 272.12 & \phantom{1}52.9 &           21.27 & 20 & 43.5 & 0.75 \\
\hline
CO ~~~~~~~~~~ 10--9 &  1151.99 & 260.24 &           304.1 & 0.10  & 19 & 13.0 & 0.65 \\
\hline
\end{tabular}
\tablefoot{
	\tablefoottext{a}{From the JPL database of molecular spectroscopy \citep{pickett98}.}
	\tablefoottext{b}{Total on $+$ off $+$ overheads integration time.}
	\tablefoottext{c}{Observed in the same setting as the main isotopologue.}}
\end{center}
\label{tab:h2o_line}
\end{table*}
}

\section{RADEX results}

\begin{figure}
\begin{center}
\includegraphics[width=0.8\columnwidth, angle=0]{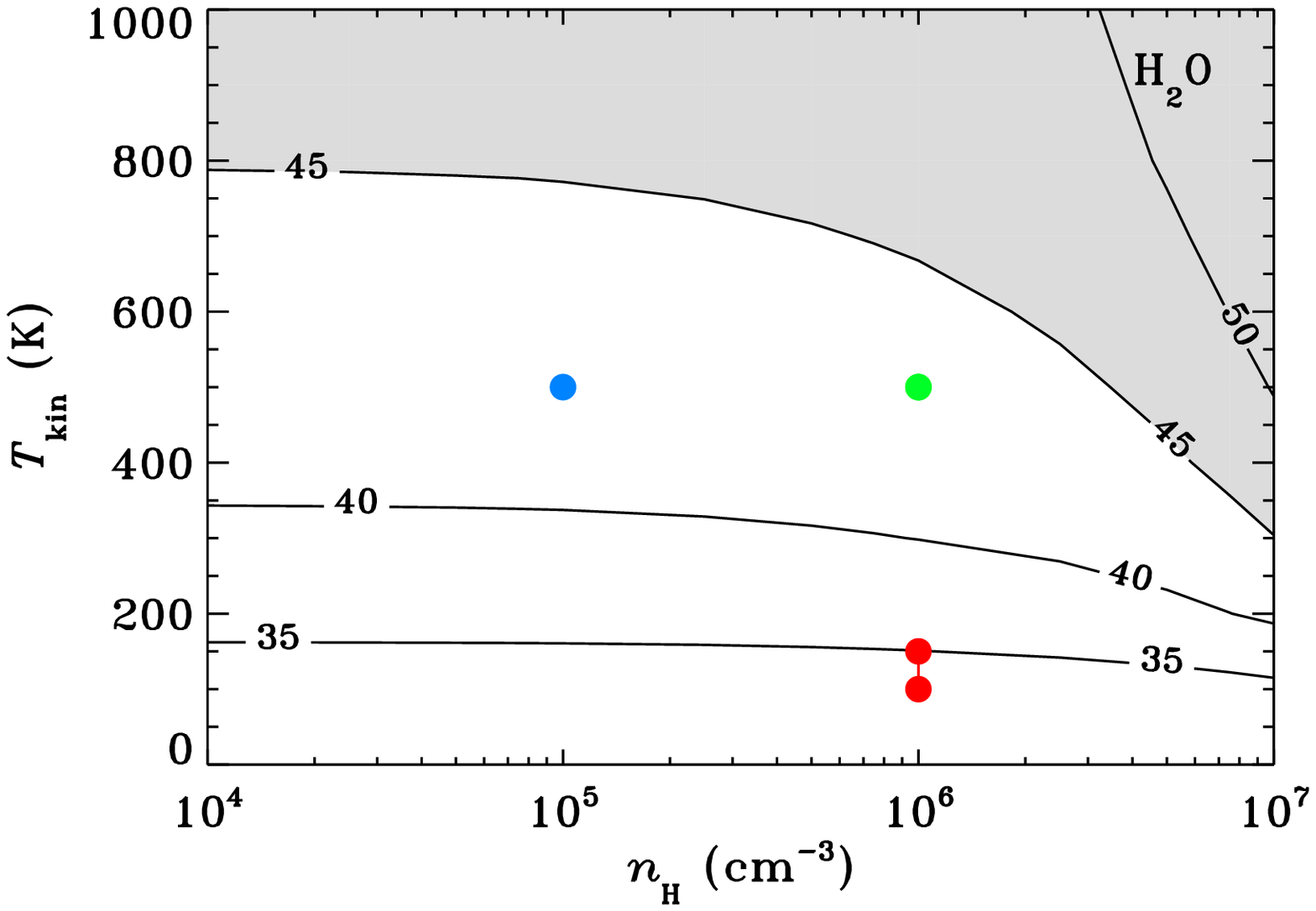}
\includegraphics[width=0.8\columnwidth, angle=0]{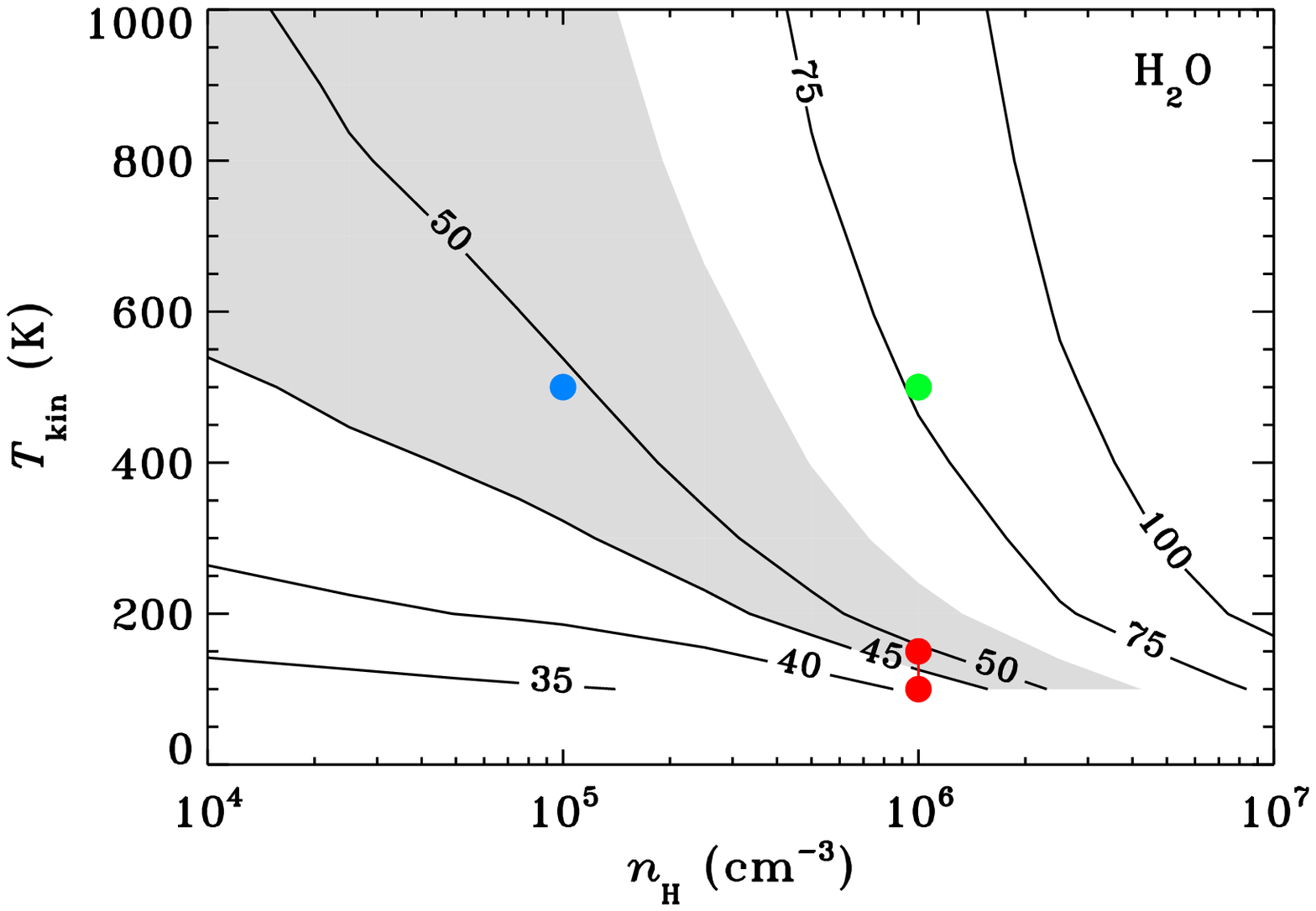}
\includegraphics[width=0.8\columnwidth, angle=0]{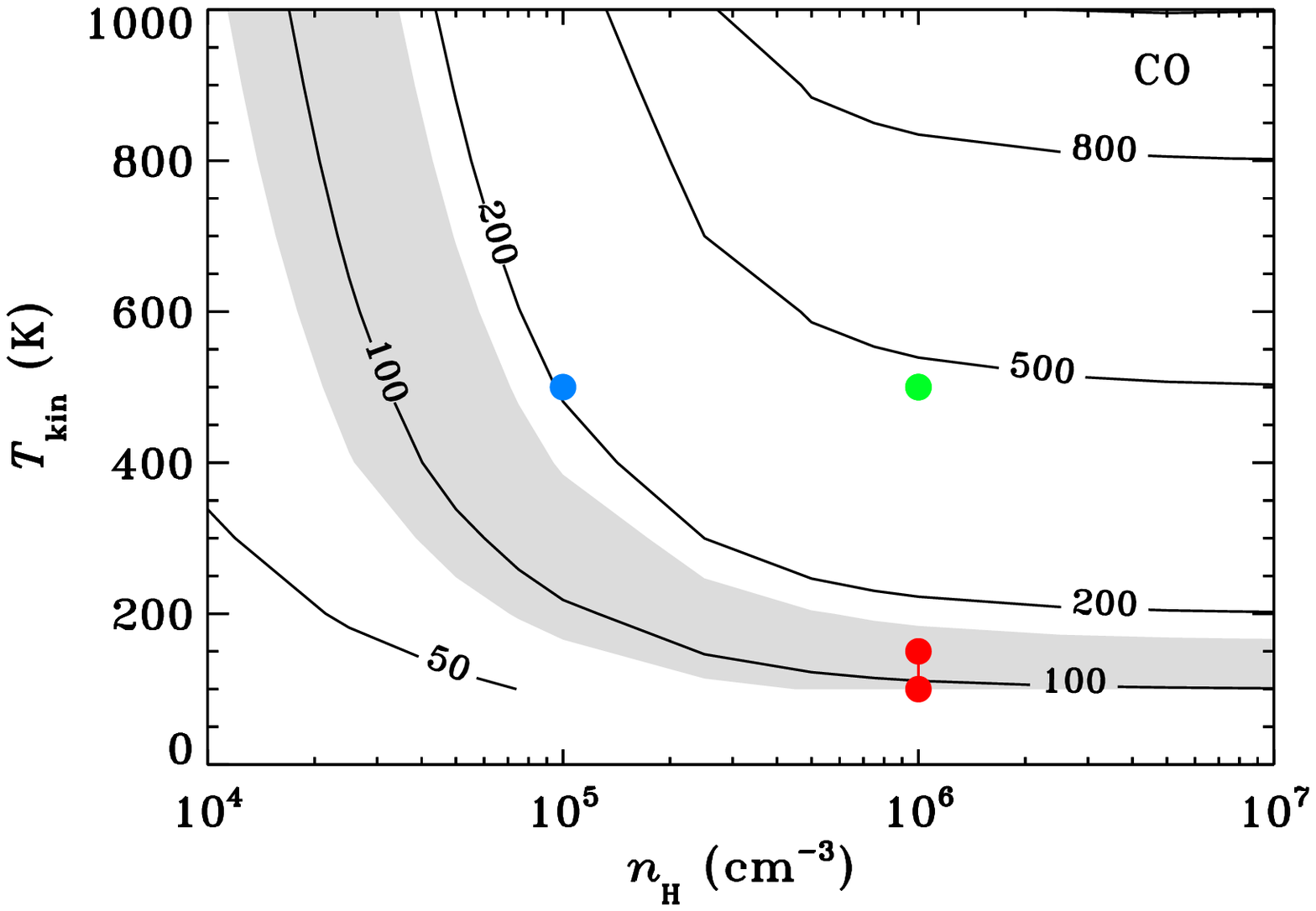}
\end{center}
\caption{$T_{\rm rot}$ of optically thin H$_2$O emission (top), optically thick H$_2$O emission ($N$=10$^{17}$ cm$^{-2}$; middle) and optically thin CO emission (bottom) determined from RADEX simulations. $T_{\rm rot}$ is calculated from the same transitions as in Sect. \ref{sec:obs}, i.e., excluding the ground-state transitions. The gray area indicates the observed values of $T_{\rm rot}$. The points indicate the different physical conditions examined; model 1 (red) has ($T$, $n$) = (150 K, 10$^6$ cm$^{-3}$) for the EHV components and (100 K, 10$^6$ cm$^{-3}$) for the broad component. Model 2 (blue) has ($T$, $n$) = (500 K, 10$^5$ cm$^{-3}$) for the EHV components, respectively. Model 3 (green) has ($T$, $n$) = (500 K, 10$^6$ cm$^{-3}$) for all components.}
\label{fig:radex_rotdiag}
\end{figure}

The non-LTE, escape-probability code RADEX was used to create rotational diagrams for H$_2$O and CO in the optically thin limit. Collisional rate coefficients are from \citet{faure07} and \citet{yang10}, respectively. The same transitions as observed were then used to calculate the rotational temperature as function of $T_{\rm kin}$ and $n_{\rm H}$. The results are shown in Fig. \ref{fig:radex_rotdiag}.

\begin{figure}
\begin{center}
\includegraphics[width=\columnwidth, angle=0]{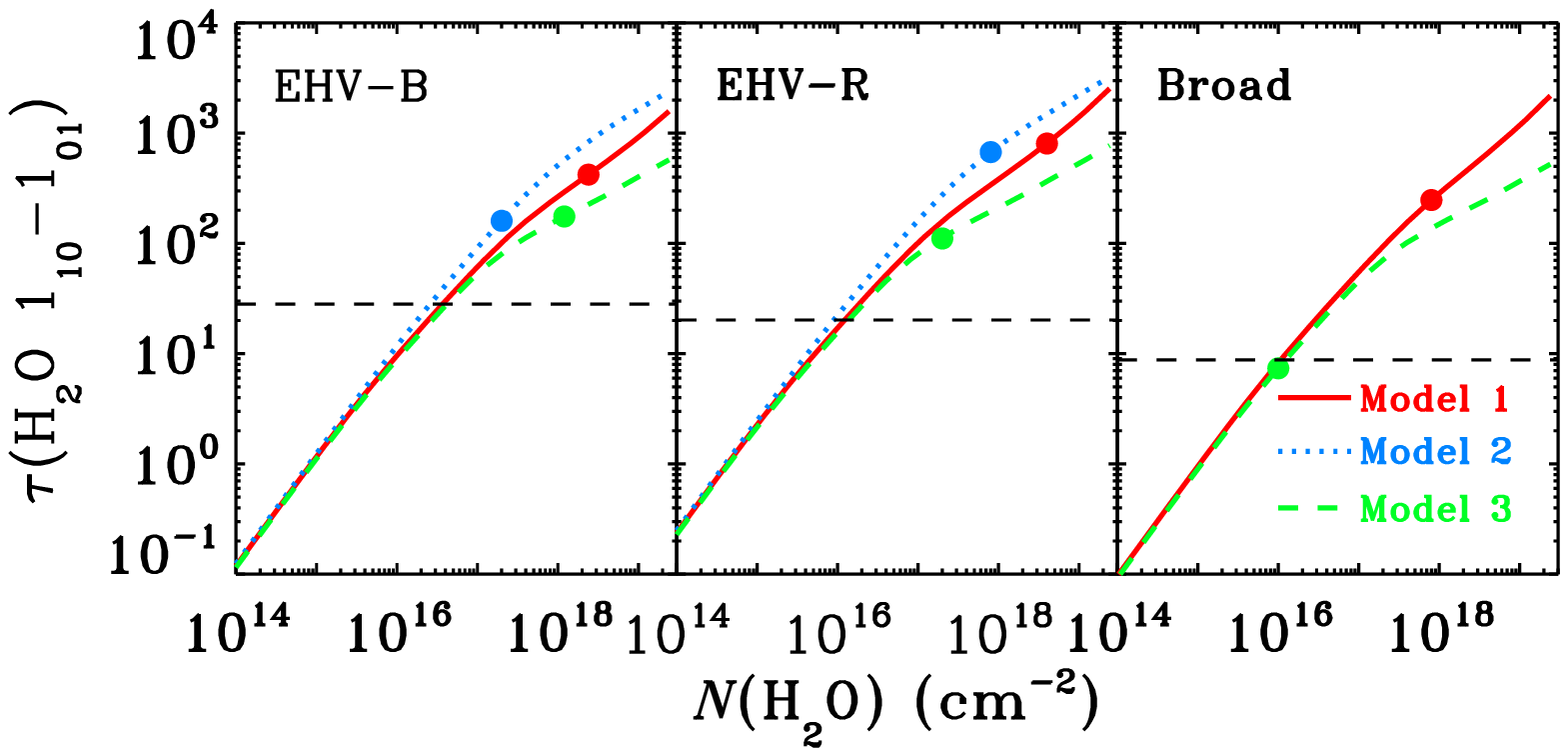}
\includegraphics[width=\columnwidth, angle=0]{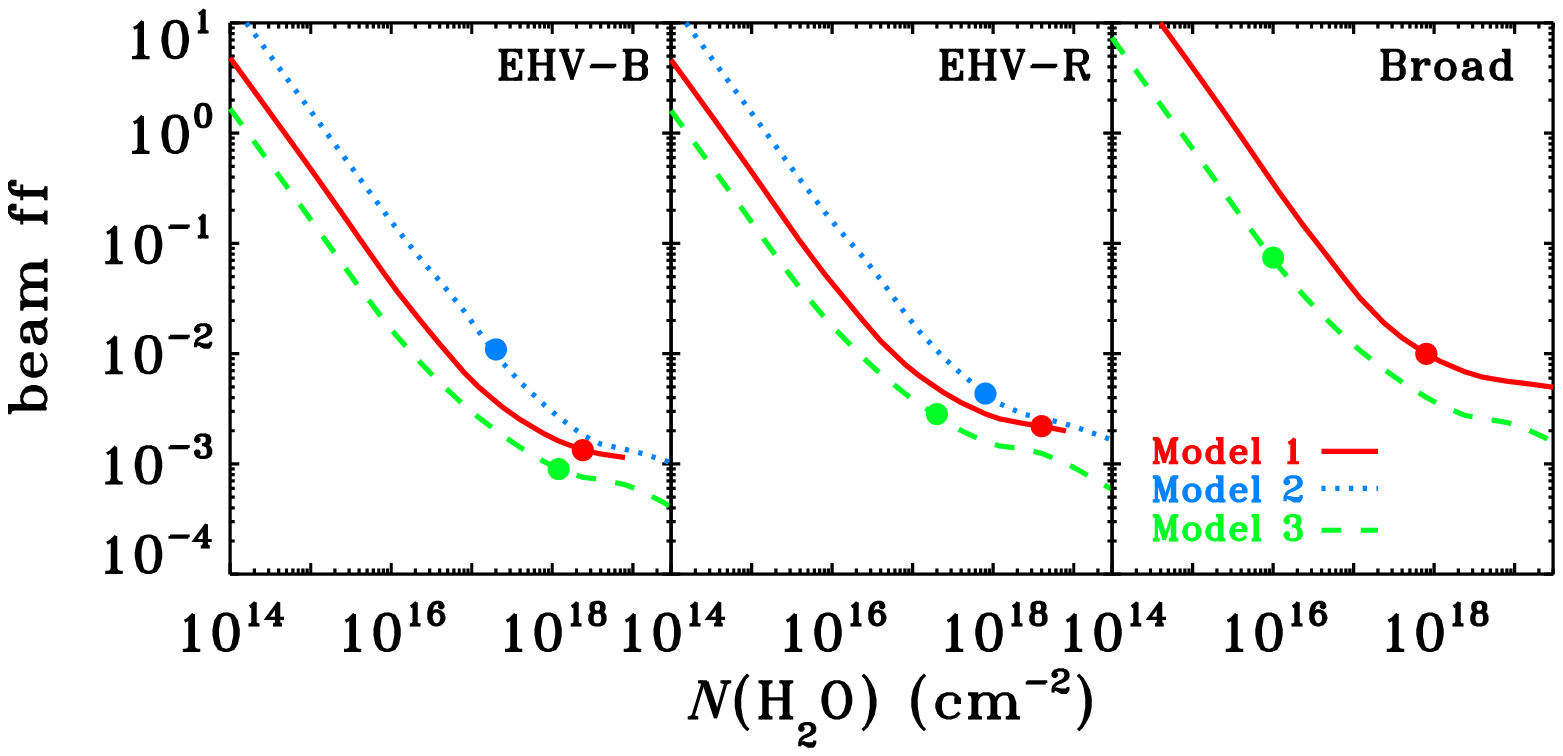}
\end{center}
\caption{(Top:) $\tau$ of the H$_2^{16}$O 1$_{10}$--1$_{01}$ line. The black dashed line corresponds to the observed value in each of the three velocity components as specified in Sect. 2. (Bottom:) Beam-filling factor for the three different components and three different models as a function of total H$_2$O column density. Best-fit results are marked with points in both plots.}
\label{fig:radex_tau}
\end{figure}

RADEX was also used to calculate the optical depth, $\tau$ of the H$_2^{16}$O 1$_{10}$--1$_{01}$ line at 557 GHz. Results are shown in Fig. \ref{fig:radex_tau}, where also the beam-filling factor is shown as a function of H$_2$O column density for the three different models.

\clearpage

\end{appendix}

\end{document}